 \documentclass[final,3p,times]{elsarticle}
\usepackage{multirow,setspace,times,amssymb,amsmath,graphicx,color,rotating,subfigure,url}
\usepackage{lineno}
\usepackage{natbib}

\journal{Physica A} 

\begin{document}

\begin{frontmatter}

\title{Empirical regularities of opening call auction in Chinese stock market}
\author[SB,RCE,SS]{Gao-Feng Gu}
\author[SB,RCE,RCSE]{Fei Ren}
\author[SB,RCE,SS]{Xiao-Hui Ni}
\author[SZSE]{Wei Chen}
\author[SB,RCE,RCSE,SS,RCFE]{Wei-Xing Zhou\corref{cor}}
\cortext[cor]{Corresponding author. Address: 130 Meilong Road, P.O.
Box 114, School of Business, East China University of Science and
Technology, Shanghai 200237, China, Phone: +86 21 64253634, Fax: +86
21 64253152.}
\ead{wxzhou@ecust.edu.cn} %

\address[SB]{School of Business, East China University of Science and Technology, Shanghai 200237, China}
\address[RCE]{Research Center for Econophysics, East China University of Science and Technology, Shanghai 200237, China}
\address[SS]{School of Science, East China University of Science and Technology, Shanghai 200237, China}
\address[RCSE]{Engineering Research Center of Process Systems Engineering (Ministry of Education), East China University of Science and Technology, Shanghai 200237, China}
\address[SZSE]{Shenzhen Stock Exchange, 5045 Shennan East Road, Shenzhen 518010, P. R. China}
\address[RCFE]{Research Center on Fictitious Economics \& Data Science, Chinese Academy of Sciences, Beijing 100080, China}

\begin{abstract}
We study the statistical regularities of opening call auction using
the ultra-high-frequency data of 22 liquid stocks traded on the
Shenzhen Stock Exchange in 2003. The distribution of the relative
price, defined as the relative difference between the order price in
opening call auction and the closing price of last trading day, is
asymmetric and that the distribution displays a sharp peak at zero
relative price and a relatively wide peak at negative relative
price. The detrended fluctuation analysis (DFA) method is adopted to
investigate the long-term memory of relative order prices. We
further study the statistical regularities of order sizes in opening
call auction, and observe a phenomenon of number preference, known
as order size clustering. The probability density function (PDF) of
order sizes could be well fitted by a $q$-Gamma function, and the
long-term memory also exists in order sizes. In addition, both the
average volume and the average number of orders decrease
exponentially with the price level away from the best bid or ask
price level in the limit-order book (LOB) established immediately
after the opening call auction, and a price clustering phenomenon is
observed.
\end{abstract}


\end{frontmatter}

\section{Introduction}

Call auction and continuous auction are two main trading mechanisms
used in order-driven financial markets. In the call auction market,
the orders arriving during the period of opening call auction are
batched and executed with a single price, i.e., the opening price
established immediately after the opening call auction, while the
continuous auction is a process of continuous matching of arriving
orders on a one-by-one basis. Much effort has been devoted to study
the market performance under these two different types of trading
mechanism. Compared with the continuous auction, the call auction
has two major advantages. Schnitzlein compared the call and
continues auction under asymmetric information in laboratory asset
market constructed based on the Kyle model \cite{Kyle-1985-Em}, and
found that the informed noise traders spend lower costs in the call
auction \cite{Schnitzlein-1996-JF}. Qualitatively similar results
have been obtained by utilizing different modeling approaches
\cite{Madhavan-1992-JF,Pagano-Roell-1996-JF}. Theissen further
confirmed that in experimental asset market incorporating
heterogeneous information the call auction provides lower execution
costs \cite{Theissen-2000-JFM}. On the other hand, the opening price
in the call auction market is closer to the true value of the asset
than the opening price in continuous market
\cite{Theissen-2000-JFM}. These two advantages are also regarded as
the goal of market construction \cite{Schreiber-Schwartz-1986-JPM},
and it has been proposed that an electronic call auction could be
incorporated into the continuous market to make it more efficient
\cite{Economides-Schwartz-1995-JPM}.

Nowadays, the call action has been widely used as the opening or
closing procedure in most electronic continuous markets. For
example, the New York Stock Exchange (NYSE), London Stock Exchange
(LSE), Euronext Paris, Frankfurt Stock Exchange (FWB), Tokyo Stock
Exchange (TSE), Hong Kong Stock Exchange (HKEX). In this paper we
mainly focus on the opening call auction in the Chinese stock
market. According to the situation of market transparency defined as
``the ability of market participants to observe the information in
trading process'', the opening call auction is divided into two
categories, i.e., close call (or blind) auction and open call
auction. Before July 1, 2006, the opening call auction of the
Shenzhen Stock Exchange was close call auction wherein the
information about submitted orders is not observable for market
participants. It is well accepted that in a sufficient large market
the transparency can improve the market efficiency
\cite{Madhavan-1996-JFI,Baruch-2005-JB}. After July 1, 2006, the
opening call auction of the Shenzhen Stock Exchange turned to be
open call auction in which the information is opened to market
participants the same as the many foreign stock exchanges, e.g. LSE,
Euronext Paris, FWB, and HKEX.

Not much work has been done to the study of the opening call auction
in the Chinese stock markets. Pan et al. proposed a theoretical
model of close call auction, and further analyzed the data of
Shanghai Stock Exchange to confirm their theoretical results that
the market should increase the transparency in the opening call
auction \cite{Pan-Liu-Liu-Wu-2004-cnSEtp}. Li et al. empirically
studied the influence of open call auction on the market volatility
in the opening of Shenzhen Stock Exchange
\cite{Xu-Li-Zeng-2007-cnJFR}. Up to now, the close call auction in
the opening of Shenzhen Stock Exchange has not been extensively
analyzed. The study of the close call auction has potential
significance for understanding the influence of transparency on
market volatility.

In this paper, we study the statistical regularities of opening call
auction for 22 liquid stocks traded on the Shenzhen Stock Exchange
in 2003 when the close call auction was adopted. The rest of the
paper is organized as follows. In Section~\ref{sec:database}, we
describe briefly the database we analyzed. Section~\ref{sec:price}
presents the statistical regularities of the order prices in opening
call auction. In Section~\ref{sec:size}, we further analyze the
order size in opening call auction. Then we study in
Section~\ref{sec:shape} the limit-order book established by the
unexecuted orders left at the end of the opening process.
Section~\ref{sec:conclusion} summarizes the results.

\section{Data sets}
\label{sec:database}

The Shenzhen Stock Exchange (SZSE) was established on December 1,
1990 and started its operations on July 3, 1991. It contains two
independent markets, A-share market and B-share market. The former
is composed of common stocks which are issued by mainland Chinese
companies. It is opened only to domestic investors, and traded in
CNY. The latter is also issued by mainland Chinese companies, while
it is traded in {\em{Hong Kong dollar}} (HKD). It was restricted to
foreign investors before February 19, 2001, and since then it has
been opened to Chinese investors as well. At the end of 2003, there
were 491 A-share stocks and 57 B-share stocks listed on the SZSE. In
the year 2003, the opening call auction is held between 9:15 am and
9:25 am, followed by the cooling periods from 9:25 am to 9:30 am,
and the continuous auction operating from 9:30 am to 11:30 am and
13:00 pm to 15:00 pm.

Our analysis is based on a database recording the order flows of 22
liquid stocks extracted from the A-share market on the SZSE in the
whole year of 2003 when the close call auction was adopted in the
opening procedure. The trading system did not show any information
about the order flows, and traders submitted orders only according
to the closing price of last trading day. The database contains the
price, size and associated time of each submitted order recorded in
the opening call with the time stamps accurate to 0.01 second. For
more details, refer to Ref. \cite{Gu-Chen-Zhou-2007-EPJB}.
Table~\ref{Tb:NOF} depicts the basic statistics of order flows in
the opening call auction for 22 stocks. Remarkably, for all the
stocks the number of sell orders $N_s$ is larger than the number of
buy orders $N_b$, and the ratio $R_N$ of $N_s$ to $N_b$ varies
within the range $[1.59, 2.80]$ with the mean value
$\overline{R}_N=2.13$. Moreover, the ratio $R_s$ of the average size
of sell orders $\langle s_s \rangle$ to the average size of buy
orders $\langle s_b \rangle$ varies within the range $[0.62, 2.35]$
with the mean value $\overline{R}_s=1.14$. The relative order size
$R_N\times R_s$ is larger than 1 for all the 22 stocks, which
indicates that the total size of sell orders is larger than the
total size of buy orders. This phenomenon was indeed observed in the
bear market during the year 2003 that the market participants were
more willing to sell.

\begin{table}[htp]
\centering \caption{Basic statistics of order flows in the opening
call auction for 22 stocks. The columns show the stock code, the
number of sell orders $N_s$, the number of buy orders $N_b$, the
ratio of $N_s$ to $N_b$ ($R_N$), the average size of sell orders
$\langle s_s \rangle$, the average size of buy orders $\langle s_b
\rangle$, the ratio of $\langle s_s \rangle$ to $\langle s_b
\rangle$ ($R_s$), the number of canceled orders $N_c$, and the
average number of daily orders $\langle N \rangle$.}
\medskip
\label{Tb:NOF} \centering
\begin{tabular}{lrrrrrrrr}
 \hline \hline
  \multicolumn{1}{c}{Code} & \multicolumn{1}{c}{$N_s$} & \multicolumn{1}{c}{$N_b$} &
  \multicolumn{1}{c}{$R_N$} & \multicolumn{1}{c}{$\langle s_s \rangle$} & \multicolumn{1}{c}{$\langle s_b \rangle$} &
  \multicolumn{1}{c}{$R_s$} & \multicolumn{1}{c}{$N_c$} & \multicolumn{1}{c}{$\langle N \rangle$}\\
  \hline
    000001 & 72,685 & 45,719 & 1.59 & 1,800 & 1,428 & 1.26 & 3,630 & 495 \\
    000002 & 48,296 & 24,098 & 2.00 & 2,896 & 2,427 & 1.19 & 290 & 303 \\
    000009 & 41,028 & 19,766 & 2.08 & 2,574 & 1,990 & 1.29 & 1,031 & 253 \\
    000012 & 18,192 & 8,368 & 2.17 & 1,843 & 1,599 & 1.15 & 393 & 112 \\
    000016 & 14,568 & 7,276 & 2.00 & 1,830 & 1,677 & 1.09 & 407 & 91 \\
    000021 & 24,387 & 13,239 & 1.84 & 1,727 & 1,435 & 1.20 & 861 & 157 \\
    000024 & 12,631 & 5,640 & 2.24 & 1,877 & 1,765 & 1.06 & 357 & 77 \\
    000027 & 35,007 & 13,435 & 2.61 & 2,386 & 2,002 & 1.19 & 349 & 203 \\
    000063 & 23,800 & 10,394 & 2.29 & 1,923 & 1,533 & 1.25 & 179 & 144 \\
    000066 & 19,860 & 9,532 & 2.08 & 1,503 & 1,174 & 1.28 & 599 & 122 \\
    000088 & 8,645 & 3,092 & 2.80 & 1,547 & 2,052 & 0.75 & 66 & 49 \\
    000089 & 19,313 & 9,519 & 2.03 & 3,996 & 6,456 & 0.62 & 147 & 122 \\
    000429 & 13,505 & 7,045 & 1.92 & 2,513 & 2,195 & 1.14 & 347 & 86 \\
    000488 & 15,104 & 9,095 & 1.66 & 4,088 & 1,738 & 2.35 & 680 & 101 \\
    000539 & 13,718 & 5,030 & 2.73 & 5,534 & 4,593 & 1.20 & 313 & 79 \\
    000541 & 12,936 & 7,034 & 1.84 & 2,305 & 2,580 & 0.89 & 258 & 83 \\
    000550 & 20,427 & 9,936 & 2.06 & 2,323 & 2,455 & 0.95 & 563 & 127 \\
    000581 & 13,531 & 5,115 & 2.65 & 1,990 & 2,303 & 0.86 & 308 & 78 \\
    000625 & 23,481 & 12,516 & 1.88 & 3,032 & 3,821 & 0.79 & 1,501 & 151 \\
    000709 & 27,200 & 13,324 & 2.04 & 3,818 & 2,881 & 1.33 & 530 & 170 \\
    000720 & 16,433 & 9,536 & 1.72 & 3,118 & 2,318 & 1.35 & 257 & 110 \\
    000778 & 22,771 & 8,858 & 2.57 & 2,415 & 2,639 & 0.92 & 575 & 132 \\
 \hline \hline
\end{tabular}
\end{table}

\section{Order price}
 \label{sec:price}

\subsection{Probability distribution of relative order prices}

In the opening call auction, we define the relative order price $x$
as the relative difference between the price of a submitted order
and the closing price of last trading day,
\begin{equation}
x(t) = \left\{
\begin{array}{llll}
 \left[p(t) - p_c \right] / p_c &&& {\rm{for~buy~orders}} \\
 \left[p_c - p(t) \right] / p_c &&& {\rm{for~sell~orders}},
\end{array}
\right. \label{Eq:x}
\end{equation}
where $p(t)$ is the price of a submitted order at time $t$, and
$p_c$ is the closing price of last trading day\footnote{Ref.
\cite{Gu-Chen-Zhou-2008b-PA} gives a wrong definition of relative
price in the opening call auction, since the virtual transaction
price is not observable for traders.}. The relative price $x$
describes the aggressiveness of a submitted order. For buy (sell)
orders, positive value of $x$ means that the trader is eager to buy
(sell) and thus place an order at a price higher (lower) than the
closing price of last trading day. Because of the 10\% price limit
trading rule in the Chinese market, the value of the relative price
$x$ is restricted to the range $[-0.1,0.1]$.

\begin{figure}[htb]
\centering
\includegraphics[width=8cm]{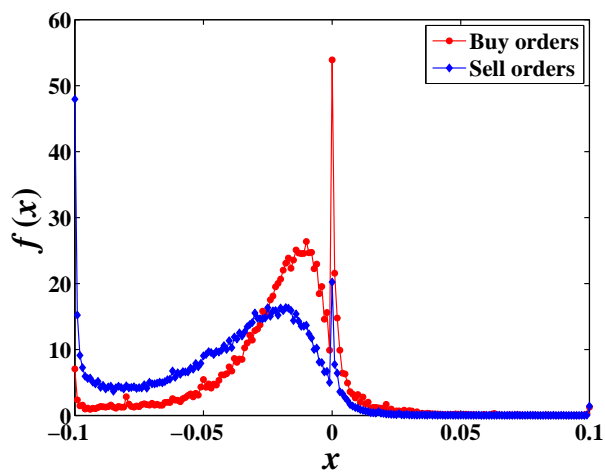}
\caption{(Color online) PDFs $f(x)$ of relative order prices $x$
using the aggregated data of 22 stocks in the opening call auction
for both buy orders and sell orders.} \label{Fig:rp}
\end{figure}

We first compute the probability distribution of relative order
prices to investigate how the traders who are only informed of the
closing price of last trading day place their orders in opening call
auction. Since all the 22 stocks have similar probability
distributions, we aggregate the data and treat all the stocks as an
ensemble. Fig.~\ref{Fig:rp} illustrates the PDFs $f(x)$ of relative
order prices for both buy orders and sell orders.

To determine the order price, the trader faces a dilemma and has to
balance the certainty of execution on one hand and the potential
benefit on the other hand. For buy orders, $f(x)$ displays a sharp
peak exhibiting a maximum at $x=0$, and displays a relatively wide
peak at negative price as shown in Fig.~\ref{Fig:rp}. There are many
traders who are eager to execute their trades and place aggressive
orders at price close to the closing price to increase the chance of
execution, while most of the traders are rational to reduce the cost
by placing their orders at prices lower than the closing price. The
$f(x)$ curve for sell orders shows similar behavior, but exhibits a
maxima at $x=-0.1$ which implies that more traders want to sell and
place orders at the highest price to minimize investment loss in the
bear market. In general, the distributions for both buy orders and
sell orders are asymmetric and skewed to the negative part. The
congregation of orders placed at $x=0,-0.1$ may suggest that the
closing price of last trading day plays an important role in order
price determination in the close call auction.

\subsection{Memory effect of relative order prices}

\begin{figure}[htb]
\centering
\includegraphics[width=8cm]{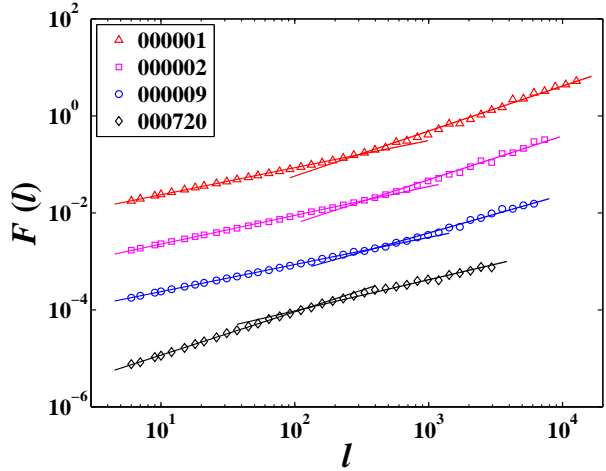}
\caption{(Color online) Fluctuation functions $F(\ell)$ of relative
order prices with respect to the time lag $\ell$ for four
representative stocks 000001, 000002, 000009 and 000720. The curves
for stocks 000002, 000009 and 000720 have been shifted vertically
for clarity.} \label{Fig:DFA_RP}
\end{figure}

Another important characteristic feature of the financial time
series is the memory effect. There are many different methods to
examine the memory effect in time series analysis. Here we use the
detrended fluctuation analysis (DFA) method
\cite{Peng-Buldyrev-Havlin-Simons-Stanley-Goldberger-1994-PRE,Kantelhardt-Bunde-Rego-Havlin-Bunde-2001-PA}
to investigate the temporal correlation of the relative order price.
$F(l)$ is expected to scale with $l$ as
\begin{equation}
   F(l)\sim l^ H,
\end{equation}
where $H$ is known as the Hurst exponent. For $H>0.5$ the time
series is long-term correlated, and for $H=0.5$ the time series is
uncorrelated. In Fig.~\ref{Fig:DFA_RP}, the fluctuation functions
$F(l)$ of relative order prices in the opening call auction for four
representative stocks are plotted.

In Fig.~\ref{Fig:DFA_RP}, we find that the fluctuation function
$F(l)$ shows two scaling regions for all the four stocks. The solid
lines are power-law fits in small scale and large scale regions
respectively. To estimate the crossover point $l_{\times}$ which
separates the two power-law regions, a simple least-squares
estimation method is applied by minimizing the square distance
between $F(l)$ and its best power-law fits in small scale and large
scale regions. For stock 000001, we obtain $l_{\times}=294$, and the
Hurst exponents are estimated to be $H_1=0.56\pm0.01$ in small scale
region and $H_2=0.93\pm0.02$ in large scale region. Using the same
method we obtain $H_1=0.59\pm0.01$ and $H_2=0.90\pm0.02$ separated
at $l_{\times}=345$ for stock 000002, $H_1=0.56\pm0.01$ and
$H_2=0.78\pm0.02$ separated at $l_{\times}=397$ for stock 000009,
and $H_1=0.89\pm0.02$ and $H_2=0.64\pm0.01$ separated at
$l_{\times}=120$ for stock 000720. Table~\ref{Tb:x_H} depicts the
Hurst exponents of relative order prices in both small scale and
large scale regions for 22 stocks in the opening call auction.

\begin{table}[htp]
\centering \caption{Hurst exponents of relative order prices for 22
stocks. The columns show the stock code, the Hurst exponent in small
scale region $H_1$, the Hurst exponent in large scale region $H_2$,
the crossover point $l_{\times}$, and the average daily Hurst
exponent $H_D$.}
\medskip
\label{Tb:x_H} \centering
\begin{tabular}{cccrc||lccrc}
 \hline \hline
  Code & $H_1$ & $H_2$ & $l_{\times}$ & $H_D$ & Code & $H_1$ & $H_2$ & $l_{\times}$ & $H_D$ \\
  \hline
    000001 & $0.56 \pm 0.01$ & $0.93 \pm 0.02$ & 294 & 0.54 & 000089 & $0.66 \pm 0.01$ & $0.70 \pm 0.02$ &  71 & 0.60 \\
    000002 & $0.59 \pm 0.01$ & $0.90 \pm 0.02$ & 345 & 0.57 & 000429 & $0.55 \pm 0.01$ & $0.77 \pm 0.04$ & 153 & 0.59 \\
    000009 & $0.56 \pm 0.01$ & $0.78 \pm 0.02$ & 397 & 0.55 & 000488 & $0.70 \pm 0.01$ & $0.79 \pm 0.03$ & 180 & 0.64 \\
    000012 & $0.58 \pm 0.01$ & $0.81 \pm 0.02$ & 209 & 0.59 & 000539 & $0.65 \pm 0.01$ & $0.86 \pm 0.03$ & 141 & 0.64 \\
    000016 & $0.55 \pm 0.01$ & $0.82 \pm 0.03$ & 162 & 0.56 & 000541 & $0.56 \pm 0.01$ & $0.79 \pm 0.03$ & 149 & 0.59 \\
    000021 & $0.58 \pm 0.01$ & $0.97 \pm 0.02$ & 253 & 0.54 & 000550 & $0.61 \pm 0.01$ & $0.98 \pm 0.04$ & 211 & 0.57 \\
    000024 & $0.64 \pm 0.01$ & $0.81 \pm 0.02$ & 136 & 0.59 & 000581 & $0.68 \pm 0.01$ & $0.84 \pm 0.02$ & 146 & 0.64 \\
    000027 & $0.60 \pm 0.01$ & $0.74 \pm 0.02$ & 221 & 0.56 & 000625 & $0.61 \pm 0.01$ & $0.95 \pm 0.01$ & 122 & 0.58 \\
    000063 & $0.62 \pm 0.01$ & $0.77 \pm 0.02$ & 235 & 0.59 & 000709 & $0.62 \pm 0.01$ & $0.77 \pm 0.01$ & 134 & 0.59 \\
    000066 & $0.58 \pm 0.01$ & $0.86 \pm 0.02$ & 162 & 0.56 & $000720~^{\star}$ & $0.89 \pm 0.02$ & $0.64 \pm 0.01$ & 120 & 0.82 \\
    000088 & $0.61 \pm 0.01$ & $0.75 \pm 0.01$ &  66 & 0.65 & 000778 & $0.59 \pm 0.01$ & $0.89 \pm 0.03$ & 237 & 0.58 \\
  \hline\hline
\end{tabular}
\end{table}

According to Table~\ref{Tb:x_H}, the Hurst exponent in small scale
region $H_1$ is slightly larger than 0.5 except for the stock 000720
marked with $\star$, which indicates that a relatively weak memory
exists in relative order prices. The crossover point $l_{\times}$
varies within the range $\left[0.5\langle N \rangle, 2\langle N
\rangle \right]$, where $\langle N \rangle$ is the average number of
daily orders for each stock as depicted in Table~\ref{Tb:NOF}. This
implies that the weak memory effect in the small scale region
persists for one or two days and then a crossover occurs. We assume
that $F(l)$ in the small scale region mainly describes the memory
effect of relative order prices within a day. To verify this, we
calculate the average daily Hurst exponent $H_D$, which is defined
as follows
\begin{equation}
H_D = \frac{1}{T_i}\sum_{j=1}^{T_i}{H_{i,j}}~, \label{Eq:H_D}
\end{equation}
where $H_{i,j}$ is the daily Hurst exponent calculated by using the
relative order prices in the opening of trading day $j$ for stock
$i$, and $T_i$ is the number of trading days for stock $i$. As shown
in Table~\ref{Tb:x_H}, $H_D$ has values similar to $H_1$, and
consequently verify our assumption. In large scale region, the Hurst
exponent $H_2$ has values apparently larger than 0.5 except for the
stock 000720. This implies that the memory effect of relative order
prices is quite strong within a period of more than one day. It is
probably due to the arrival of important news or events which
affects investors' trading behavior and makes them continuously buy
or sell within a period of several days or weeks.

\section{Order size}
 \label{sec:size}

\subsection{Number preference of order sizes}

\begin{figure}[htb]
\centering
\includegraphics[width=4.5cm]{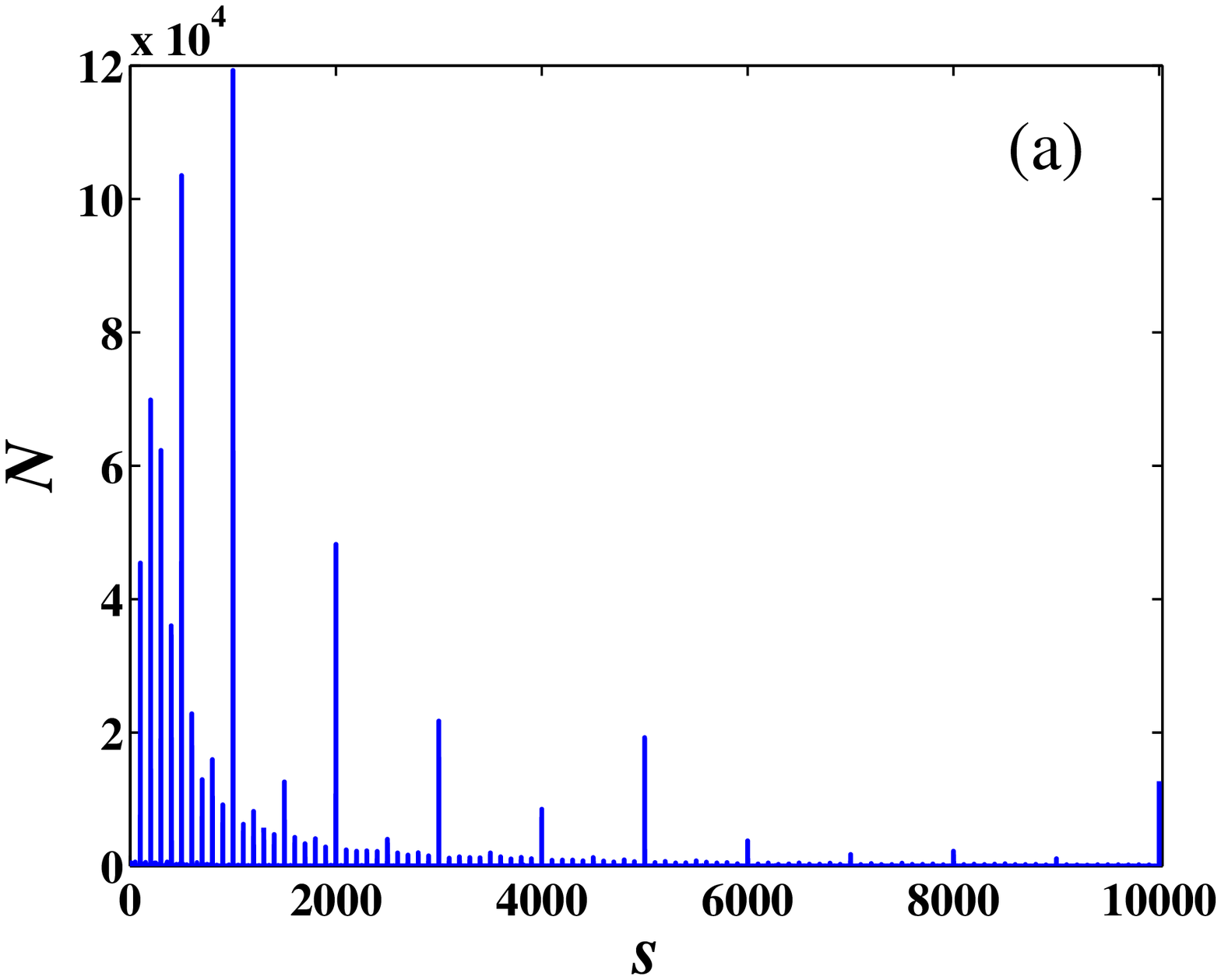}
\includegraphics[width=4.5cm]{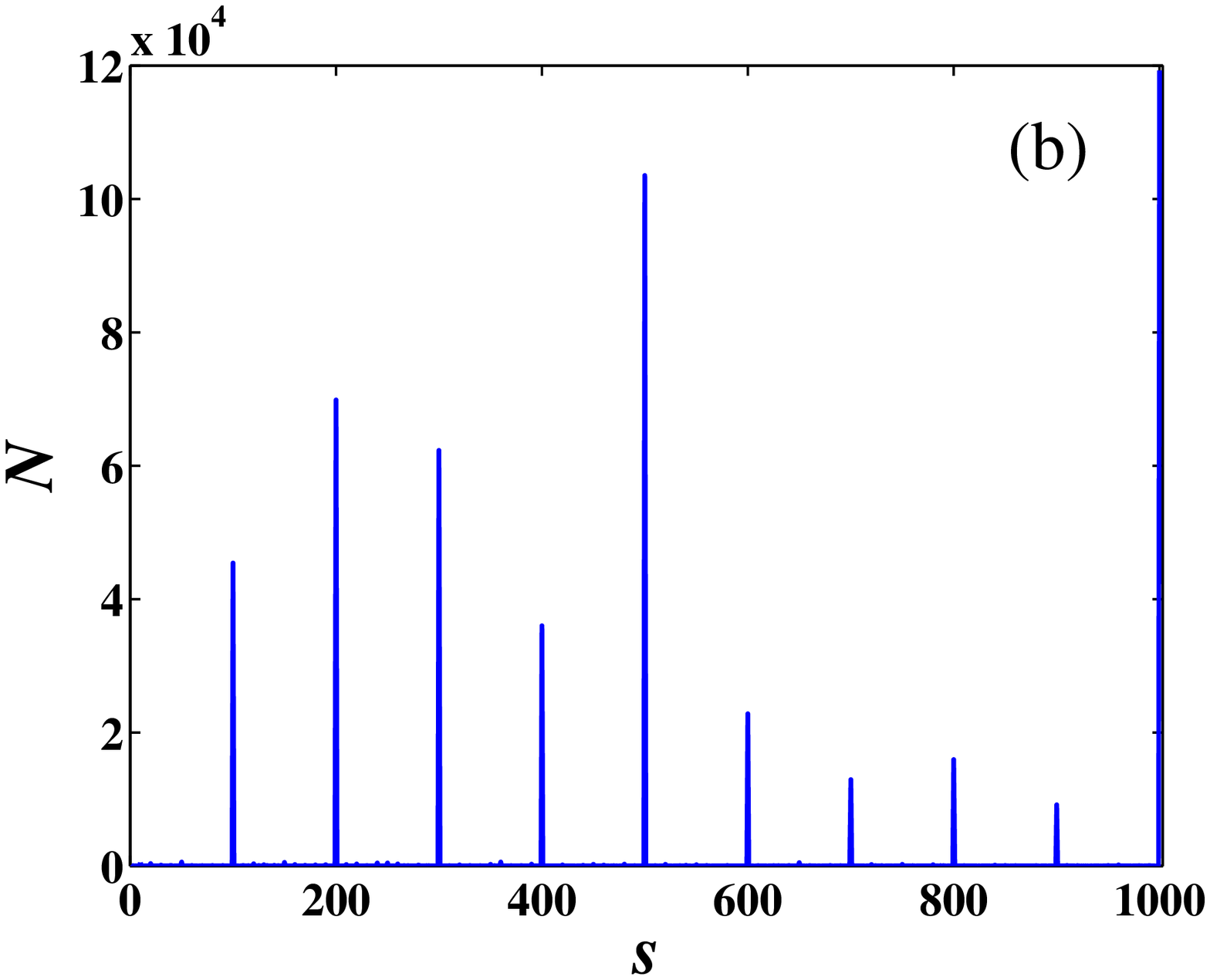}
\includegraphics[width=4.5cm]{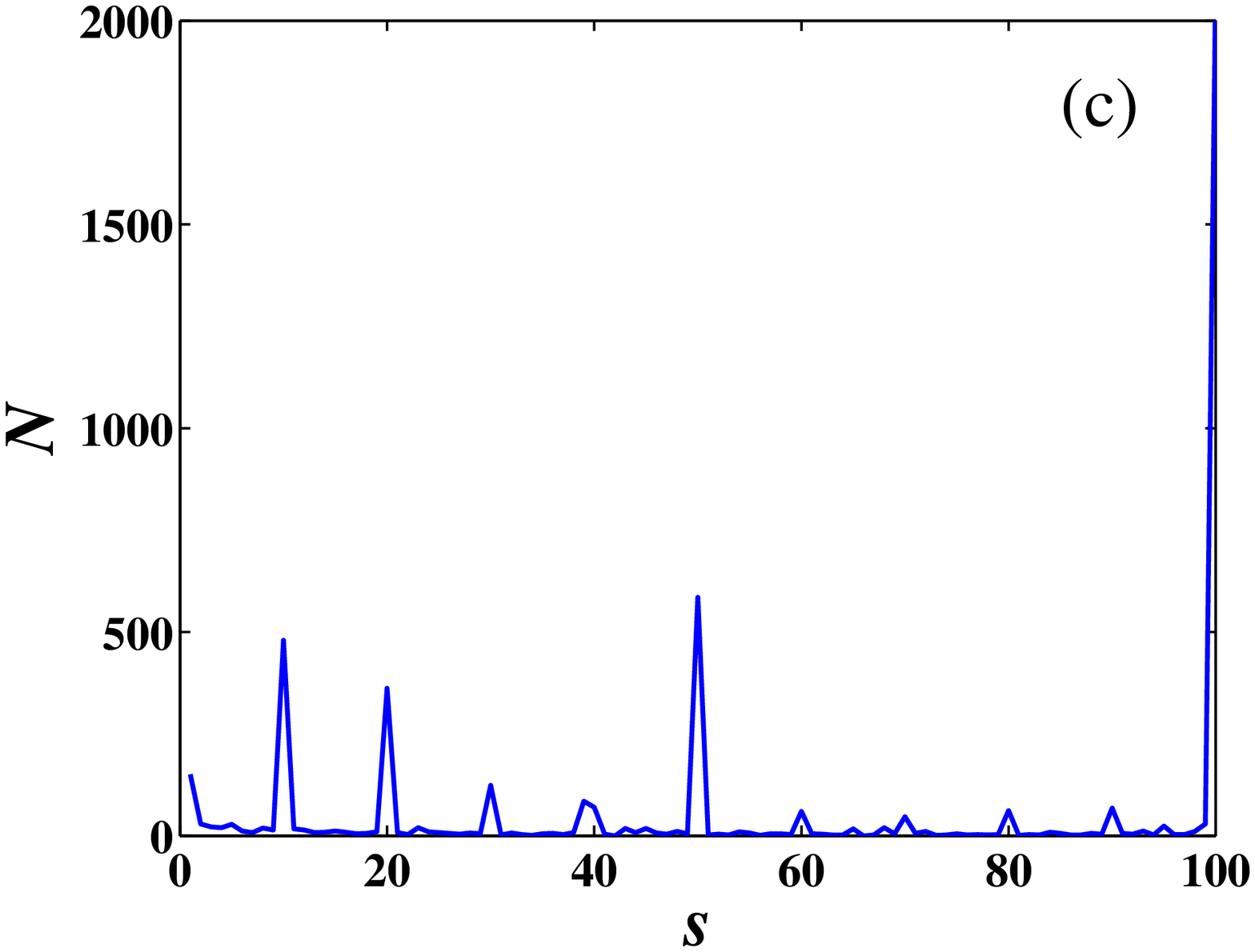}
\caption{(Color online) Histogram plots of order sizes in the
opening call auction at three different scales: (a) $s\in
\left[1,10^4 \right]$, (b) $s\in \left[1,10^3 \right]$, and (c)
$s\in \left[1,10^2 \right]$.} \label{Fig:clustering_Size}
\end{figure}

We then study another ingredient of submitted orders, i.e., order
size. The order size plays an important role in the dynamics of
price formation
\cite{Karpoff-1987-JFQA,Chan-Fong-2000-JFE,Lillo-Farmer-Mantegna-2003-Nature,Lim-Coggins-2005-QF,Zhou-2007-XXX},
as a well-known adage says ``it takes volume to move stock prices.''
In Fig.~\ref{Fig:clustering_Size}, we plot the number of submitted
orders which have the same size as a function of the order size $s$
using the aggregated data of the 22 stocks.

In Fig.~\ref{Fig:clustering_Size}, one observes that there exists
several layers of spikes in the histogram plots of order sizes at
different scales. It may be caused by the number preference
phenomenon of order sizes similar to the trade size clustering
phenomenon which universally exists in finance markets
\cite{Moulton-2005-JFE,Hodrick-Moulton-2007-XXXX,Alexander-Peterson-2007-JFE,Mu-Chen-Kertesz-Zhou-2009-EPJB}.
We name it as order size clustering phenomenon. The traders usually
prefer to place orders with size following the formula
$s=c\times10^k$, where $c$ and $k$ are integers.
Fig.~\ref{Fig:clustering_Size} (a) shows the histogram plot of order
sizes in a large scale region $s\in \left[1,10^4 \right]$, and a
layer of spikes is displayed with spikes located at $s=c\times10^3$,
$c=1,2,\cdots,10$. Fig.~\ref{Fig:clustering_Size} (b) shows the
histogram plot of order sizes in a smaller scale region $s\in
\left[1,10^3 \right]$, and another layer of spikes is displayed with
spikes located at $s=c\times10^2$, $c=1,2,\cdots,10$. The histogram
plot of order sizes in a small scale region $s\in \left[1,10^2
\right]$ is shown in Fig.~\ref{Fig:clustering_Size} (c), and a
similar layer of spikes is observed with spikes located at
$s=10,20,\cdots,100$. It mainly describes the regularity of sell
orders, because in Chinese stock markets buy orders must be in a
board lot of 100 shares or their multiples, while sales of stocks
with less than 100 shares could be made in one order. This
phenomenon for call auction is the same as that for continuous
double auction \cite{Mu-Chen-Kertesz-Zhou-2009-EPJB}.

\subsection{Probability distribution of order sizes}

\begin{figure}[htb]
\centering
\includegraphics[width=6cm]{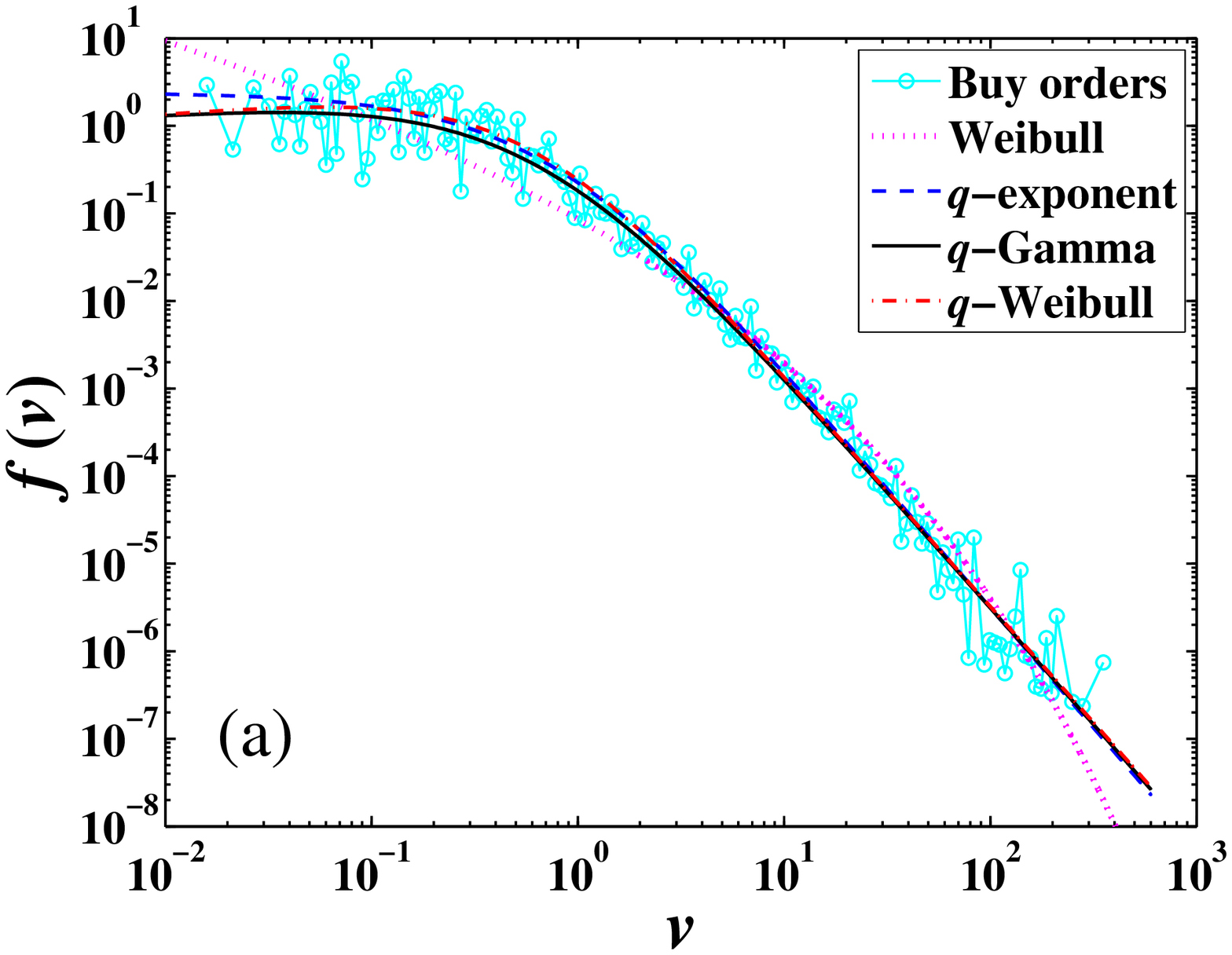}
\includegraphics[width=6cm]{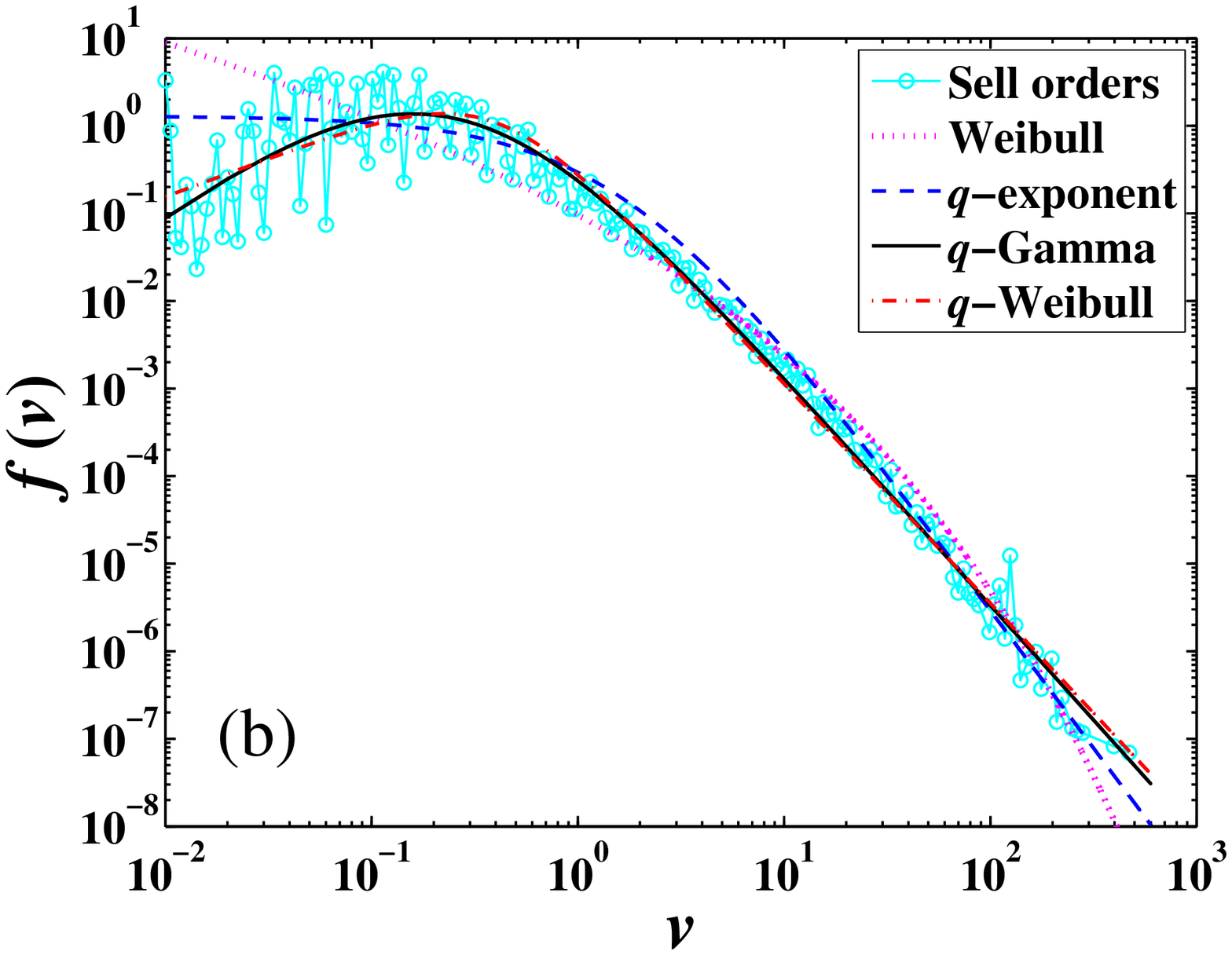}
\caption{(Color online) PDFs $f(v)$ of normalized order sizes $v$
for (a) buy orders and (b) sell orders. Dotted line, dashed line,
solid line, and dash-dotted line are fitted curves of Weibull
distribution, $q$-exponential distribution, $q$-Gamma distribution,
and $q$-Weibull distribution, respectively.} \label{Fig:PDF_Size}
\end{figure}

We normalize the order size by dividing its average value for each
stock as $v=s/\langle s\rangle$, thus the normalized order size is
in units of the average order size. We treat the 22 stocks as an
ensemble and aggregate the data. The empirical PDFs $f(v)$ of
normalized order sizes for both buy orders and sell orders are shown
in Fig.~\ref{Fig:PDF_Size}.

Four types of distribution functions are applied to fit the
empirical PDFs using the least-squares method. We first use the
Weibull distribution function
\cite{Cohen-1965-TCHN,Goda-Fukunaga-1986-JMatS,Picoli-Mendes-Malacarne-2003-PA}
which follows
\begin{equation}
f_{W}(v)= \frac{\beta}{\theta}
\left(\frac{v}{\theta}\right)^{\beta-1}{\rm{exp}}\left[-\left(\frac{v}{\theta}\right)^{\beta}\right]~,
 \label{Eq:W}
\end{equation}
where $\theta > 0$ and $\beta > 0$. The parameters fitted from the
empirical PDFs are estimated to be $\theta = 0.103$ and $\beta =
0.317$ with the error sum of squares $sse = 0.935$ for buy orders,
and $\theta = 0.149$ and $\beta = 0.332$ with $sse = 1.482$ for sell
orders.

A $q$-exponential distribution function
\cite{Mu-Chen-Kertesz-Zhou-2009-EPJB,Burr-1942-AMS,Tsallis-1988-JSP,Queiros-2005-EPL,Nadarajah-Kotz-2007-PA}
is defined as
\begin{equation}
f_{qE}(v)=
\frac{1}{\theta}\left[1-(1-q){\frac{v}{\theta}}\right]^{\frac{q}{1-q}}~,
 \label{Eq:qE}
\end{equation}
where $\theta > 0$ and $q > 1$. We then use this function to fit the
empirical PDFs, and obtain the parameters $\theta = 0.408$ and $q =
1.577$ with $sse = 0.725$ for buy orders, and $\theta = 0.771$ and
$q = 1.462$ with $sse = 1.04$ for sell orders.

We also use a $q$-Gamma distribution function
\cite{Mu-Chen-Kertesz-Zhou-2009-EPJB,Queiros-2005-EPL,Nadarajah-Kotz-2007-PA,Tsallis-Anteneodo-Borland-Osorio-2003-PA,Queiros-Moyano-deSouza-Tsallis-2007-EPJB}
defined as
\begin{equation}
\begin{aligned}
f_{qG}(v)= \frac{1}{z} \left(\frac{v}{\theta}\right)^{\beta}
\left[1-(1-q){\frac{v}{\theta}}\right]^{\frac{1}{1-q}}~,\\
z=\int_{0}^{\infty}\left(\frac{v}{\theta}\right)^{\beta}\left[1-(1-q){\frac{v}{\theta}}\right]^{\frac{1}{1-q}}{\rm{d}}v,
 \label{Eq:qG}
\end{aligned}
\end{equation}
where $\theta > 0$, $\beta > 0$, $q > 1$, and $z$ is a normalized
constant. The parameters are estimated to be $\theta = 0.216$,
$\beta = 0.155$ and $q = 1.354$ with $sse = 0.703$ for buy orders,
and $\theta = 0.062$, $\beta = 1.585$ and $q = 1.237$ with $sse =
0.739$ for sell orders.

Finally, we use a $q$-Weibull distribution function
\cite{Picoli-Mendes-Malacarne-2003-PA,Nadarajah-Kotz-2007-PA} ,
which has a form
\begin{equation}
f_{qW}(v)=
\frac{(2-q)\beta}{\theta}\left(\frac{v}{\theta}\right)^{\beta-1}\left[1-(1-q)\left(\frac{v}{\theta}\right)^{\beta}\right]^{\frac{1}{1-q}}~,
 \label{Eq:qW}
\end{equation}
where $\theta > 0$, $\beta > 0$ and $1 < q < 2$. We obtain the
parameters $\theta = 0.264$, $\beta = 1.161$ and $q = 1.414$ with
$sse = 0.720$ for buy orders, and $\theta = 0.282$, $\beta = 1.887$
and $q = 1.557$ with $sse = 0.791$ for sell orders.

In Fig.~\ref{Fig:PDF_Size}, the fitted curves of these four types of
distributions are also illustrated. It is obviously that the Weibull
distribution fits the empirical distribution worse than the $q$-type
distributions for both buy orders and sell orders, and it is further
manifested by the fact that $sse$ of the Weibull distribution shows
values larger than that of the $q$-type distributions. The $q$-Gamma
and $q$-Weibull distributions show very similar behaviors, and they
can better approximate the empirical distribution than $q$-exponent
distribution especially for small order size. Among all these four
types of distributions, the $q$-Gamma distribution can best fit the
empirical distribution, since it has the smallest value of $sse$.

\subsection{Memory effect of order sizes}

\begin{figure}[htb]
\centering
\includegraphics[width=8cm]{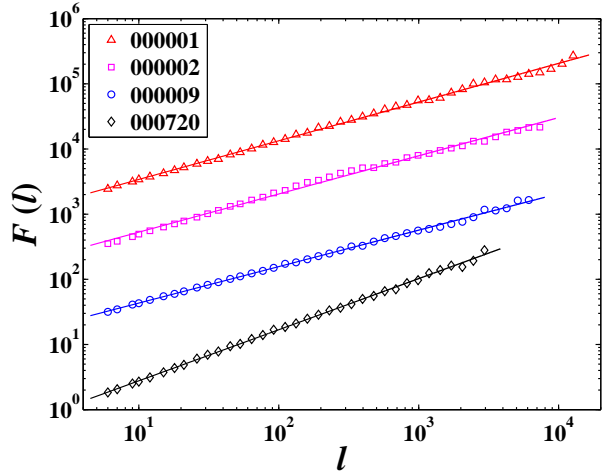}
\caption{(Color online) Fluctuation functions $F(l)$ of order sizes
for four representative stocks 000001, 000002, 000009 and 000720.
The curves for stocks 000002, 000009 and 000720 have been shifted
vertically for clarity.} \label{Fig:DFA_Size}
\end{figure}

We also apply the DFA method to investigate the temporal correlation
of order sizes. Fig.~\ref{Fig:DFA_Size} illustrates the fluctuation
functions $F(l)$ of order sizes in the opening call auction for four
representative stocks. A power law is clearly observed in the whole
region of the order size. The Hurst exponent is estimated to be
$H=0.59\pm0.01$ for stock 000001, $H=0.58\pm0.01$ for stock 000002,
$H=0.56\pm0.01$ for stock 000009, and $H=0.78\pm0.01$ for stock
000720. Table~\ref{Tb:s_H} depicts the Hurst exponents of order
sizes in the opening call auction for 22 stocks. The Hurst exponents
for all the stocks are larger than 0.5, which indicates that the
long-term memory also exists in order sizes.

\begin{table}[htp]
\centering \caption{Hurst exponents of order sizes for 22 stocks.
The columns show the stock code and its corresponding Hurst
exponent.}
\medskip
\label{Tb:s_H} \centering
\begin{tabular}{cc||cc||cc}
 \hline \hline
  Code & $H$ & Code & $H$ & Code & $H$ \\
  \hline
    000001 & $0.59 \pm 0.01$ & 000063 & $0.63 \pm 0.01$ & 000550 & $0.57 \pm 0.01$ \\
    000002 & $0.58 \pm 0.01$ & 000066 & $0.59 \pm 0.01$ & 000581 & $0.63 \pm 0.01$ \\
    000009 & $0.56 \pm 0.01$ & 000088 & $0.58 \pm 0.01$ & 000625 & $0.68 \pm 0.01$ \\
    000012 & $0.63 \pm 0.01$ & 000089 & $0.65 \pm 0.01$ & 000709 & $0.68 \pm 0.01$ \\
    000016 & $0.55 \pm 0.01$ & 000429 & $0.55 \pm 0.01$ & 000720 & $0.78 \pm 0.01$ \\
    000021 & $0.58 \pm 0.01$ & 000488 & $0.65 \pm 0.01$ & 000778 & $0.55 \pm 0.01$ \\
    000024 & $0.64 \pm 0.01$ & 000539 & $0.69 \pm 0.02$ &        & \\
    000027 & $0.61 \pm 0.01$ & 000541 & $0.59 \pm 0.02$ &        & \\
 \hline\hline
\end{tabular}
\end{table}

\section{Averaged shape of limit-order book (LOB)}
 \label{sec:shape}

\begin{figure}[htb]
\centering
\includegraphics[width=6cm]{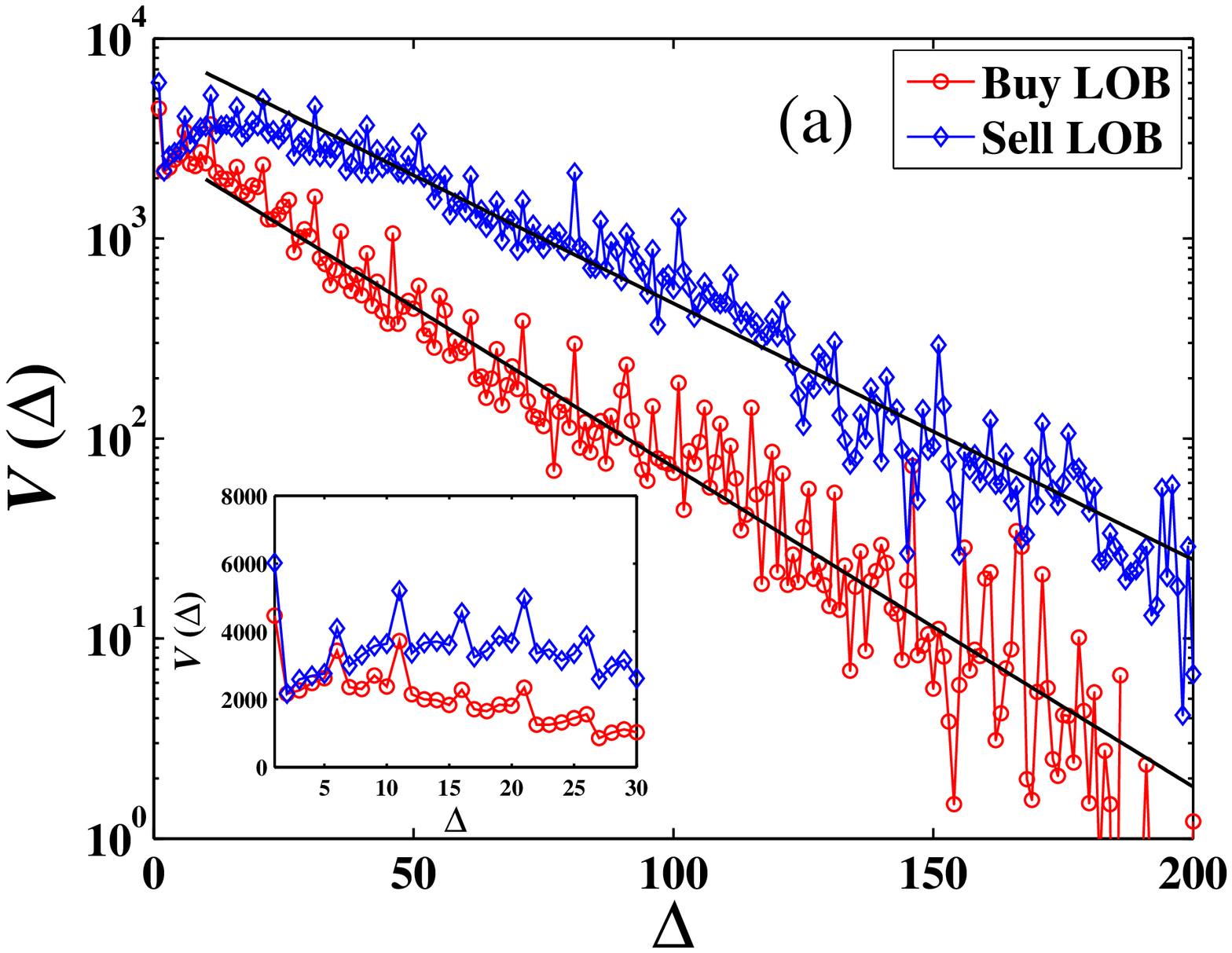}
\includegraphics[width=6cm]{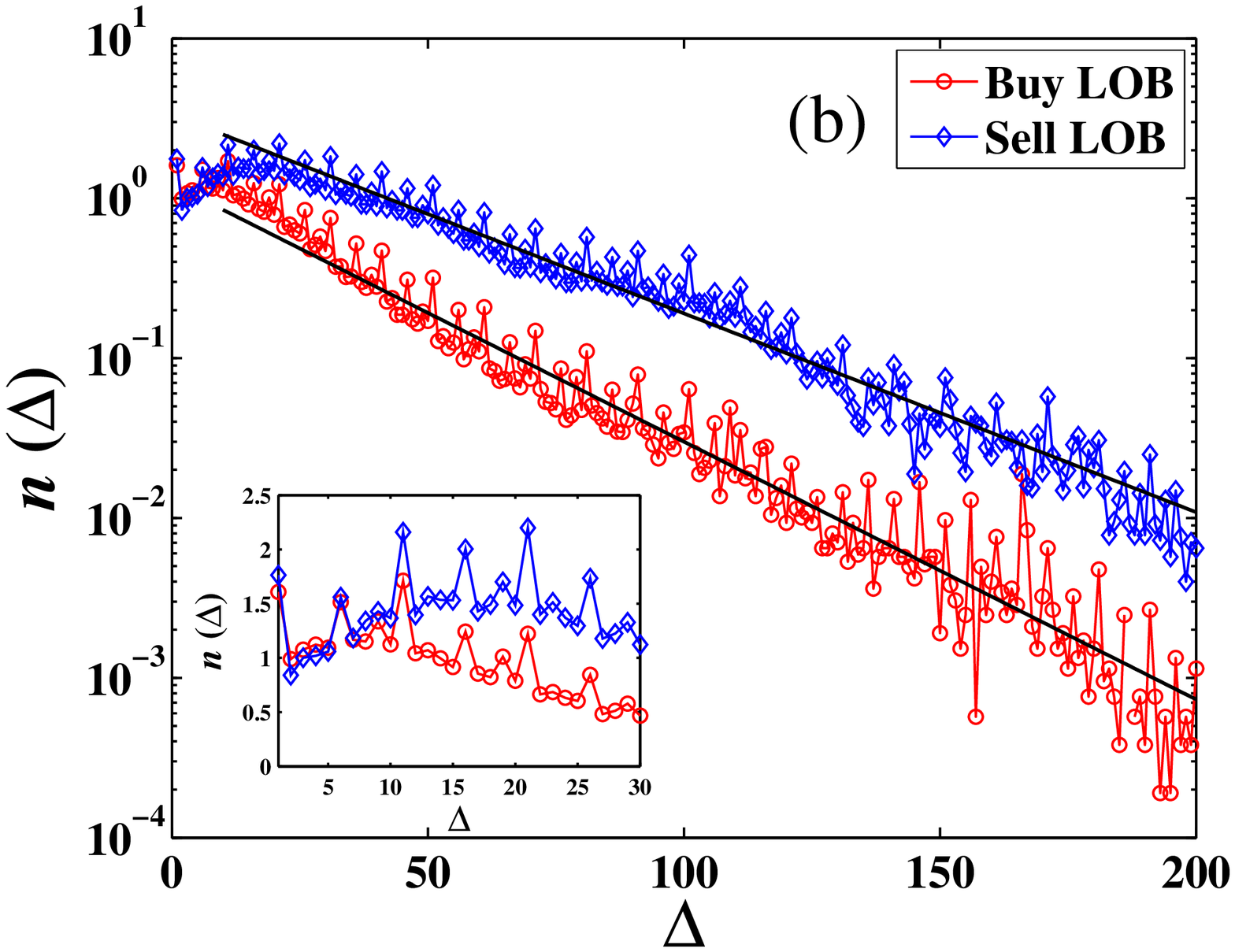}
\caption{(Color online) (a) Average order size $V(\Delta)$ and (b)
Average number of orders $n(\Delta)$ as a function of the price
level $\Delta$ for both buy LOB and sell LOB.} \label{Fig:shape}
\end{figure}

At the end of the opening procedure of each trading day, a LOB is
established based upon the unexecuted orders left in the open call
auction. Price levels are discrete in the LOB. In the Chinese stock
market, the tick size $u$ defined as the difference between two
neighbored price levels is 0.01 CNY. As with this tick size, we
define the price level $\Delta$ as the distance between the order we
considered and the best bid or ask,
\begin{equation}
\Delta = \left\{
\begin{array}{llll}
 (p_b - p)/u + 1 &&& {\rm{for~buy~orders}} \\
 (p - p_a)/u + 1 &&& {\rm{for~sell~orders}},
\end{array}
\right.
 \label{Eq:Delta}
\end{equation}
where $p$ is the order price in the LOB, and $p_b$ and $p_a$ are the
best bid price and best ask price respectively. According to
Eq.~(\ref{Eq:Delta}), $\Delta = 1$ stands for the position at the
best bid (ask) in the buy (sell) LOB. We define $V_{b,i}(\Delta,t)$
($V_{s,i}(\Delta,t)$) as the order size placed at price level
$\Delta$ in the buy (sell) LOB at day $t$ for stock $i$. We
aggregate the data of 22 stocks, and calculate the average order
size as follows
\begin{equation}
V_{\{b,s\}}(\Delta) =
 \frac{1}{MT_i}\sum_{i = 1}^M\sum_{t = 1}^{T_i} V_{\{b,s\},i}(\Delta,t)~,
 \label{Eq:V_bs}
\end{equation}
where $T_i$ is the number of trading days for stock $i$ and $M$ is
the number of stocks analyzed. In our study, $M=22$. We also
consider the number of orders placed at each price level in the LOB.
Denote the variable $n_{b,i}(\Delta,t)$ ($n_{s,i}(\Delta,t)$) as the
number of orders placed at price level $\Delta$, and the average
number of orders $n_b$ ($n_s$) is defined as
\begin{equation}
n_{\{b,s\}}(\Delta) =
 \frac{1}{MT_i}\sum_{i = 1}^M\sum_{t = 1}^{T_i}
 n_{\{b,s\},i}(\Delta,t)~.
 \label{Eq:n_bs}
\end{equation}
We plot the average order size $V(\Delta)$ and the average number of
orders $n(\Delta)$ for both buy LOB and sell LOB in
Fig.~\ref{Fig:shape}.

In Fig.~\ref{Fig:shape} (a), $V(\Delta)$ follows a linear decrease
in linear-log coordinates which implies it may decrease
exponentially for both buy LOB and sell LOB
\begin{equation}
 V_{b,s}(\Delta) \sim e^{-\beta_{b,s} \Delta}~.
 \label{Eq:V_Exp}
\end{equation}
We obtain that $\beta_b=0.0373\pm0.001$ for the buy LOB and
$\beta_s=0.0295\pm0.001$ for the sell LOB. It is clear that the
curve for buy LOB decreases more rapidly than the curve for sell LOB
and the average order size of sell orders is larger than that of the
sell orders especially for large $\Delta$. This is consistent with
the fact that the Chinese stock market in 2003 was bearish and more
market participants tended to sell their shares. Take a more careful
look at $V(\Delta)$, one observes that there are more orders placed
at the price level $\Delta = 1$ than other price levels in both buy
LOB and sell LOB as illustrated in the inset plot of
Fig.~\ref{Fig:shape} (a). This is distinctly different from the
shape of LOB in the continuous auction
\cite{Gu-Chen-Zhou-2008c-PA,Bouchaud-Mezard-Potters-2002-QF,Potters-Bouchaud-2003-PA}.
In addition, a series of periodic peaks are displayed at $\Delta =
5n+1$ for $n = 0,1,2,\ldots$, which might be related to the trader's
irrational preference of some numbers like 5, 10 or their multiples
in order price placement, known as the price clustering phenomenon
\cite{Gu-Chen-Zhou-2008c-PA,Niederhoffer-1965-OR,Niederhoffer-1966-JB,Ball-Torous-Tschoegl-1985-JFM}.
Some large peaks presented at higher price levels in sell LOB may
due to the burst of total order size of sell orders at $x=-0.1$ as
illustrated in Fig.~\ref{Fig:rp}.

As shown in Fig.~\ref{Fig:shape} (b), $n(\Delta)$ shows a similar
exponentially decreasing tendency as
\begin{equation}
n_{b,s}(\Delta) \sim e^{-\gamma_{b,s} \Delta}~,
 \label{Eq:n_Exp}
\end{equation}
and we obtain that $\gamma_b=0.0371\pm0.001$ for the buy LOB and
$\gamma_s=0.0286\pm0.001$ for the sell LOB. For both buy LOB and
sell LOB, the values of $\beta$ are very close to the values of
$\gamma$. This may suggests that the average order size placed at
each price level is independent of the order price, and crucially
depends on the number of orders.

\section{Conclusion}
\label{sec:conclusion}

Based on the order flow data of 22 liquid stocks traded on the
Shenzhen Stock Exchange in 2003, we analyze the statistical
regularities of the relative order price, the order size in the
opening call auction and the LOB shape immediately after it. The PDF
of the relative order price $x$ is asymmetric, and displays a sharp
peak at $x = 0$ and a relatively wide peak at negative $x$. the
congregation of orders placed at $x=0,-0.1$ implies the importance
of closing price of last trading day in order price determination in
the close call auction. We use the DFA method to investigate the
memory effect of relative order prices, and find the fluctuate
function $F(l)$ shows two scaling regions. $F(l)$ in the small scale
region describes the relatively weak memory effect of relative order
prices within one day, while $F(l)$ in the large scale region
describes quite persistent memory effect within a period of several
days or weeks.

We then analyze the order size in the opening call auction. Layers
of spikes are clearly observed in the histogram plot of order sizes,
which may be caused by the number preference phenomenon existing in
the order submission. We further apply four types of distribution
functions to fit the PDF of normalized order sizes and find that
$q$-Gamma distribution gives a better fit than Weibull distribution,
$q$-exponential distribution and $q$-Weibull distribution. The Hurst
exponent of order sizes is larger than 0.5, which indicates the
long-term memory also exists in order sizes. Considering the shape
of the LOB established immediately after the opening call auction,
we find that both the average order size and the average number of
orders follow exponential decays with similar exponents.

\bigskip
{\textbf{Acknowledgments:}}

This work was partially supported by the Shanghai Educational
Development Foundation (2008CG37 and 2008SG29), the National Natural
Science Foundation of China (70501011 and 70502007), and the Program
for New Century Excellent Talents in University (NCET-07-0288).

\bibliography{E:/papers/Auxiliary/Bibliography}

\begin{thebibliography}{10}
\expandafter\ifx\csname url\endcsname\relax
  \def\url#1{\texttt{#1}}\fi
\expandafter\ifx\csname urlprefix\endcsname\relax\def\urlprefix{URL }\fi

\bibitem{Kyle-1985-Em}
A.~S. Kyle, {Continuous auctions and insider trading}, Econometrica 53 (1985)
  1315--1335.

\bibitem{Schnitzlein-1996-JF}
C.~R. Schnitzlein, {Call and continuous trading mechanisms under asymmetric
  information: an experimental investigation}, J. Financ. 51 (1996) 613--636.

\bibitem{Madhavan-1992-JF}
A.~Madhavan, {Trading mechanisms in securities markets}, J. Financ. 2 (1992)
  607--641.

\bibitem{Pagano-Roell-1996-JF}
M.~Pagano, A.~R{\"o}ell, {Transparency and liquidity: A comparison of auction
  and dealer markets with informed trading}, J. Financ. 51 (1996) 579--611.

\bibitem{Theissen-2000-JFM}
E.~Theissen, {Market structure, informational efficiency and liquidity: An
  experimental comparison of auction and dealer markets}, J. Financ. Markets 3
  (2000) 333--363.

\bibitem{Schreiber-Schwartz-1986-JPM}
P.~S. Schreiber, R.~A. Schwartz, {Price discovery in securities markets}, J.
  Portfolio Management 12 (1986) 43--48.

\bibitem{Economides-Schwartz-1995-JPM}
N.~Economides, R.~A. Schwartz, {Electronic call market trading}, J. Portfolio
  Management 21 (1995) 10--18.

\bibitem{Madhavan-1996-JFI}
A.~Madhavan, {Security prices and market transparency}, Journal of Financial
  Intermediation 5 (1996) 255--283.

\bibitem{Baruch-2005-JB}
S.~Baruch, {Who benefits from an open limit-order book?}, J. Business 78 (2005)
  1267--1306.

\bibitem{Pan-Liu-Liu-Wu-2004-cnSEtp}
D.~Pan, T.~Liu, H.~L. Liu, C.~F. Wu, {Trading strategies in blind call auction:
  Models and empirical analysis of Shanghai stock exchang}, Sys. Engin. Theory
  Prac. (in Chinese) 24(1) (2004) 1--10.

\bibitem{Xu-Li-Zeng-2007-cnJFR}
X.~C. Xu, P.~Li, Y.~Zeng, {Empirical analysis of the influence of open call
  auction on volatility in Chinese stock markets}, J. Financ. Res. (in Chinese)
  07(A) (2007) 151--164.

\bibitem{Gu-Chen-Zhou-2007-EPJB}
G.-F. Gu, W.~Chen, W.-X. Zhou, {Quantifying bid-ask spreads in the Chinese
  stock market using limit-order book data: Intraday pattern, probability
  distribution, long memory, and multifractal nature}, Eur. Phys. J. B 57
  (2007) 81--87.

\bibitem{Gu-Chen-Zhou-2008b-PA}
G.-F. Gu, W.~Chen, W.-X. Zhou, {Empirical regularities of order placement in
  the Chinese stock market}, Physica A 387 (2008) 3173--3182.

\bibitem{Peng-Buldyrev-Havlin-Simons-Stanley-Goldberger-1994-PRE}
C.-K. Peng, S.~V. Buldyrev, S.~Havlin, M.~Simons, H.~E. Stanley, A.~L.
  Goldberger, {Mosaic organization of DNA nucleotides}, Phys. Rev. E 49 (1994)
  1685--1689.

\bibitem{Kantelhardt-Bunde-Rego-Havlin-Bunde-2001-PA}
J.~W. Kantelhardt, E.~Koscielny-Bunde, H.~H.~A. Rego, S.~Havlin, A.~Bunde,
  {Detecting long-range correlations with detrended fluctuation analysis},
  Physica A 295 (2001) 441--454.

\bibitem{Karpoff-1987-JFQA}
J.~M. Karpoff, {The relation between price changes and trading volume: A
  survey}, J. Financ. Quart. Anal. 22 (1987) 109--126.

\bibitem{Chan-Fong-2000-JFE}
K.~Chan, W.~M. Fong, {Trade size, order imbalance, and the volatility-volume
  relation}, J. Financ. Econ. 57 (2000) 247--273.

\bibitem{Lillo-Farmer-Mantegna-2003-Nature}
F.~Lillo, J.~D. Farmer, R.~Mantegna, {Master curve for price impact function},
  Nature 421 (2003) 129--130.

\bibitem{Lim-Coggins-2005-QF}
M.~Lim, R.~Coggins, {The immediate price impact of trades on the Australian
  Stock Exchange}, Quant. Financ. 5 (2005) 365--377.

\bibitem{Zhou-2007-XXX}
W.-X. Zhou, {Universal price impact functions of individual trades in an
  order-driven market}, http://arxiv.org/abs/0708.3198v2 (2007).

\bibitem{Moulton-2005-JFE}
P.~C. Moulton, {You can't always get what you want: Trade-size clustering and
  quantity choice in liquidity}, J. Financ. Econ. 78 (2005) 89--119.

\bibitem{Hodrick-Moulton-2007-XXXX}
L.~S. Hodrick, P.~C. Moulton, {Liquidity: considerations of a portfolio
  manager}, http://www.bnet.fordham.edu/pmoulton/Liq08292007.pdf (2007).

\bibitem{Alexander-Peterson-2007-JFE}
G.~J. Alexander, M.~A. Peterson, {An analysis of trade-size clustering and its
  relation to stealth trading}, J. Financ. Econ. 84 (2007) 435--471.

\bibitem{Mu-Chen-Kertesz-Zhou-2009-EPJB}
G.-H. Mu, W.~Chen, J.~Kert{\'e}sz, W.-X. Zhou, {Preferred numbers and the
  distributions of trade sizes and trading volumes in the Chinese stock
  market}, Eur. Phys. J. B 68 (2009) 145--152.

\bibitem{Cohen-1965-TCHN}
A.~C. Cohen, {Maximum likelihood estimation in the Weibull distribution based
  on complete and on censored samples}, Technometrics 7 (1965) 579--588.

\bibitem{Goda-Fukunaga-1986-JMatS}
K.~Goda, H.~Fukunaga, {The evaluation of the strength distribution of silicon
  carbide and alumina fibres by a multi-modal Weibull distribution}, J.
  Materials Sci. 21 (1986) 4475--4480.

\bibitem{Picoli-Mendes-Malacarne-2003-PA}
S.~Picoli~Jr., R.~S. Mendes, L.~C. Malacarne, {q-exponential, Weibull, and
  q-Weibull distributions: an empirical analysis}, Physica A 324 (2003)
  678--688.

\bibitem{Burr-1942-AMS}
I.~W. Burr, {Cumulative frequency functions}, Ann. Math. Stat. 13 (1942)
  215--232.

\bibitem{Tsallis-1988-JSP}
C.~Tsallis, {Possible generalization of Boltzmann-Gibbs statistics}, J. Stat.
  Phys. 52 (1988) 479--487.

\bibitem{Queiros-2005-EPL}
S.~M.~D. Queiros, {On the emergence of a generalised Gamma distribution:
  Application to traded volume in financial markets}, Europhys. Lett. 71 (2005)
  339--345.

\bibitem{Nadarajah-Kotz-2007-PA}
S.~Nadarajah, S.~Kotz, {On the $q$-type distributions}, Physica A 377 (2007)
  465--468.

\bibitem{Tsallis-Anteneodo-Borland-Osorio-2003-PA}
C.~Tsallis, C.~Anteneodo, L.~Borland, R.~Osorio, {Nonextensive statistical
  mechanics and economics}, Physica A 324 (2003) 89--100.

\bibitem{Queiros-Moyano-deSouza-Tsallis-2007-EPJB}
S.~M.~D. Queiros, L.~G. Moyano, J.~de~Souza, C.~Tsallis, {A nonextensive
  approach to the dynamics of financial observables}, Eur. Phys. J. B 55 (2007)
  161--167.

\bibitem{Gu-Chen-Zhou-2008c-PA}
G.-F. Gu, W.~Chen, W.-X. Zhou, {Empirical shape function of limit-order books
  in the Chinese stock market}, Physica A 387 (2008) 5182--5188.

\bibitem{Bouchaud-Mezard-Potters-2002-QF}
J.-P. Bouchaud, M.~M{\'e}zard, M.~Potters, {Statistical properties of stock
  order books: empirical results and models}, Quant. Financ. 2 (2002) 251--256.

\bibitem{Potters-Bouchaud-2003-PA}
M.~Potters, J.-P. Bouchaud, {More statistical properties of order books and
  price impact}, Physica A 324 (2003) 133--140.

\bibitem{Niederhoffer-1965-OR}
V.~Niederhoffer, {Clustering of stock prices}, Oper. Res. 13 (1965) 258--265.

\bibitem{Niederhoffer-1966-JB}
V.~Niederhoffer, {A new look at clustering of stock prices}, J. Business 39
  (1966) 309--313.

\bibitem{Ball-Torous-Tschoegl-1985-JFM}
C.~A. Ball, W.~N. Torous, A.~E. Tschoegl, {The degree of price resolution: the
  case of the gold market}, J. Fut. Markets 5 (1985) 29--43.

\end{thebibliography}

\end{document}